\documentclass[useAMS,usenatbib,twocolumn,fleqn]{mn2e}
\usepackage{natbib, graphics, epsfig}

\usepackage{times}
\usepackage{amsmath}
\usepackage{amssymb}
\usepackage{textcomp}
\usepackage{verbatim}
\newcommand{\sm}{\small}

\title[Constraining Self-Interacting Dark Matter with the Milky Way's dwarf spheroidals]
      {Constraining Self-Interacting Dark Matter with the Milky Way's dwarf spheroidals}
      \author[J. Zavala, M. Vogelsberger \& M.~G. Walker] {\parbox{18.5cm}{
          Jes\'us Zavala$^{1,2}$\thanks{CITA National Fellow, e-mail: jzavalaf@uwaterloo.ca},          
          Mark Vogelsberger$^{3}$\thanks{Hubble Fellow} and
          Matthew G. Walker$^3$\thanks{Hubble Fellow},
        }\vspace{0.3cm}\\
        $^{1}$Perimeter Institute for Theoretical Physics, 31 Caroline St. N., Waterloo, ON, N2L 2Y5, Canada\\
        $^{2}$Department of Physics and Astronomy, University of Waterloo, Waterloo, Ontario, N2L 3G1, Canada\\
        $^3$Harvard-Smithsonian Center for Astrophysics, 60 Garden Street, Cambridge, MA 02138, USA}

\begin{document}
\date{Accepted ???. Received ???; in original form ???}

\pagerange{\pageref{firstpage}--\pageref{lastpage}} \pubyear{2012}

\maketitle

\label{firstpage}

\begin{abstract}
Self-Interacting Dark Matter is an attractive alternative to the Cold Dark Matter paradigm only if it is able to 
substantially reduce the central densities of dwarf-size haloes while 
keeping the densities and shapes of cluster-size haloes within current constraints. Given the seemingly stringent nature
of the latter, it was thought for nearly a decade that Self-Interacting Dark Matter would be viable only if the cross section for self-scattering 
was strongly velocity-dependent. However, it has recently been suggested that a constant cross section per unit mass of
${\sigma_T/m\!\sim\!0.1\,{\rm cm}^2\,{\rm g}^{-1}}$ is sufficient to accomplish the desired effect.
We explicitly investigate this claim using high resolution cosmological simulations of a Milky-Way size
halo and find that, similarly to the Cold Dark Matter case, such cross section produces a population of massive subhaloes that is inconsistent with the 
kinematics of the classical dwarf spheroidals, in particular with the inferred slopes of the mass profiles of Fornax and Sculptor. 
This problem is resolved if ${\sigma_T/m\!\sim\!1\,{\rm cm}^2\,{\rm g}^{-1}}$ at the dwarf spheroidal scales. Since this value is likely inconsistent
with the halo shapes of several clusters, our results leave only a small window open for a velocity-independent Self-Interacting Dark Matter  model to work as a
distinct alternative to Cold Dark Matter.
\end{abstract}

\begin{keywords}
cosmology: dark matter -- galaxies: halos -- methods: numerical 
\end{keywords}

\section{Introduction}

It is now clear that observations of dark matter dominated systems such as low-mass (dwarf) and low surface 
brightness (LSB) galaxies favour the presence of dark matter cores of ${\mathcal{O}(1\,{\rm kpc})}$
(e.g. \citealt{Moore1994}, \citealt{Kuzio2008}, \citealt{deBlok2010}, \citealt{Walker2011}, \citealt{Amorisco2012}).
These observations are a challenge for the Cold Dark Matter (CDM) 
paradigm where dark matter haloes are predicted to have density cusps, an imprint of the 
collisionless nature of CDM (the {\it core-cusp problem}). In a possibly related issue, it has been 
pointed out recently that the dark satellites of Milky-Way (MW) size halo simulations are too dense to be consistent 
with the kinematics of the MW dwarf spheroidals (dSphs) (the {\it too big to fail problem};
\citealt{Boylan2011,Boylan2012}). 

Although these are significant challenges to the CDM model, their solution could naturally lie in our incomplete 
understanding of the complex process of galaxy formation. Even though the internal dynamics of dwarfs is 
dominated by dark matter today, it is conceivable that earlier episodes of star formation and subsequent 
gas removal by supernova feedback might have been violent enough to modify the initial cuspy dark matter 
distribution into a cored one \citep[e.g.][]{Navarro1996,Pontzen2012}. Current hydrodynamical 
simulations have shown that such a mechanism is able to create large cores in intermediate-mass galaxies 
\citep{Governato2010,Governato2012}, and, once tidal stripping is taken into account, it might solve 
the too big to fail problem as well \citep{Brooks2012}. It is not clear, however, if such
episodes of large gas blow-outs are consistent with the star formation histories 
and stellar properties of LSBs and dwarfs \citep[e.g.][]{Kuzio2011, Boylan2012,Penarrubia2012}.

Given the large uncertainties regarding whether baryon physics can reconcile the CDM model with observations
of dwarf galaxies, it is prudent to consider the alternative, which is to question the fundamental CDM hypotheses, namely, the 
collisionless and cold nature of CDM particles. This alternative is additionally encouraged by the
null detection of several experiments that are pursuing the discovery of the favoured CDM particles, and whose 
sensitivity is reaching the natural values for the interaction cross sections of the
particle physics models that predict them (e.g. Supersymmetry, \citealt{Abazajian2012,Xenon2012}).

An exciting possibility is that of self-interacting dark matter (SIDM) originally introduced over a
decade ago by \citet{SpergelSteinhardt2000}. Self-scattering between dark matter particles is 
a feature of present hidden-sector dark matter models that predict the existence of new gauge bosons. The presence of
these bosons is invoked to enhance the annihilation and/or self-scattering of dark matter particles to 
explain a number of puzzling observations (the {\it Sommerfeld enhancement}, e.g. 
\citealt{Arkani-Hamed2009}, \citealt{Buckley2010}).
Collisional dark matter is constrained by the requirements from different astrophysical observations, 
such as the ellipsoidal shape of haloes, the avoidance of subhalo evaporation in galaxy clusters, and the 
avoidance of the gravothermal catastrophe \citep[e.g.][]{Miralda2002,Gnedin2001,Firmani2001}. 
The original excitement caused by SIDM died off by the apparently strong constraints on the scattering cross section
set particularly by X-ray and lensing observations of clusters in the analysis by \citet{Miralda2002}: 
${\sigma_T/m\!\leq\!0.02\,{\rm cm}^2\,{\rm g}^{-1}}$;
such a low cross section would have no relevant impact for the dynamics of galaxies at the ${\mathcal{O}(1\,{\rm kpc})}$
scale. \citet{Peter2012} have revised this constraint and found that it was overestimated by over an
order of magnitude suggesting that a current constraint is of ${\mathcal{O}(0.1\,{\rm cm}^2\,{\rm g}^{-1}}$).

SIDM is clearly a viable model if the cross section depends on 
the relative velocity in such a way that dark matter behaves as a collisional fluid in dwarfs, 
and is essentially collisionless at the scale of clusters.
Although this idea was phenomenologically proposed \citep{Yoshida2000} and explored with 
cosmological simulations \citep{Colin2002} a decade ago, its theoretical support has come up only recently
\citep[e.g.][]{Ackerman2009,Feng2009,Feng2010,Buckley2010,Loeb2011,vanden2012,Tulin2012}. Moreover, earlier 
simulations lacked the resolution needed to reliably
explore the sub-${\rm kpc}$ region of dwarf-size haloes. It was only until recently
that it was explicitly shown that theoretically motivated velocity-dependent SIDM (vdSIDM) models 
produce core sizes consistent with those found in MW dSphs, and also solve
the emergent too big to fail problem \citep[][hereafter VZL]{VZL2012}. 

On the other hand, \citet{Rocha2012} have suggested that a 
velocity-dependent cross section is not essential since a SIDM model with a constant 
${\sigma_T/m\!=\!0.1\,{\rm cm}^2\,{\rm g}^{-1}}$ (allowed by cluster constraints) is also consistent with the inner structure of dSphs. 
This seems to contradict earlier estimates made by \citet{Yoshida2000} who 
suggested that the average number of collisions per particle in the central core, $n_{\rm core}$, scales as the cube root 
of the halo mass. Since for this cross section, ${n_{\rm core}\!\sim\!2}$ for a cluster-size halo (within ${\!\sim\!10\%}$ 
of the virial radius, see their Fig. 2), $n_{\rm core}$ would thus be suppressed
by a factor of ${\mathcal{O}(100)}$ in dwarf-size haloes, resulting in cores that are too 
small. This scaling is however imprecise since for a halo of virial mass $M$:
\begin{equation}
n_{\rm core}\!\sim\!\left<\rho_{\rm core}\right>\left(\frac{\sigma_T}{m}\right)\left<\sigma_{\rm vel}\right>t_{\rm age}
\propto\Delta_{\rm core}\left(\frac{\sigma_T}{m}\right)M^{1/3}t_{\rm age},
\end{equation}
where ${\left<\rho_{\rm core}\right>}$ and ${\left<\sigma_{\rm vel}\right>}$ are the average density and velocity 
dispersion of dark matter particles, $t_{\rm age}$ is the formation time, and $\Delta_{\rm core}$
is proportional to the density contrast relative to the background density; 
all of these quantities are defined within the core of the halo. 
Thus, the higher concentrations and larger formation times of dwarf-size haloes reduce the $M^{1/3}$
dependence. It is important to remark that both, \citet{Yoshida2000} and \citet{Rocha2012}, 
extrapolated the regimes they could directly simulate
to the regime of dwarfs. In this Letter, we resolve 
this issue by proving explicitly that a constant scattering cross-section of 
${\sigma_T/m\!=\!0.1\,{\rm cm}^2\,{\rm g}^{-1}}$ is not able 
to create ${\mathcal{O}(1\,{\rm kpc})}$ cores in the dark subhaloes where the MW dSphs are expected to live; it deviates only slightly 
from the CDM predictions. Unless baryonic processes are invoked, the range of interesting constant $\sigma_T/m$ values is 
thus very narrow: $0.1\,{\rm cm}^2\,{\rm g}^{-1}\!<\!\sigma_T/m\!<\!1\,{\rm cm}^2\,{\rm g}^{-1}$.

\section{Simulations and Results}

\begin{figure}
\centering
\includegraphics[width=0.475\textwidth]{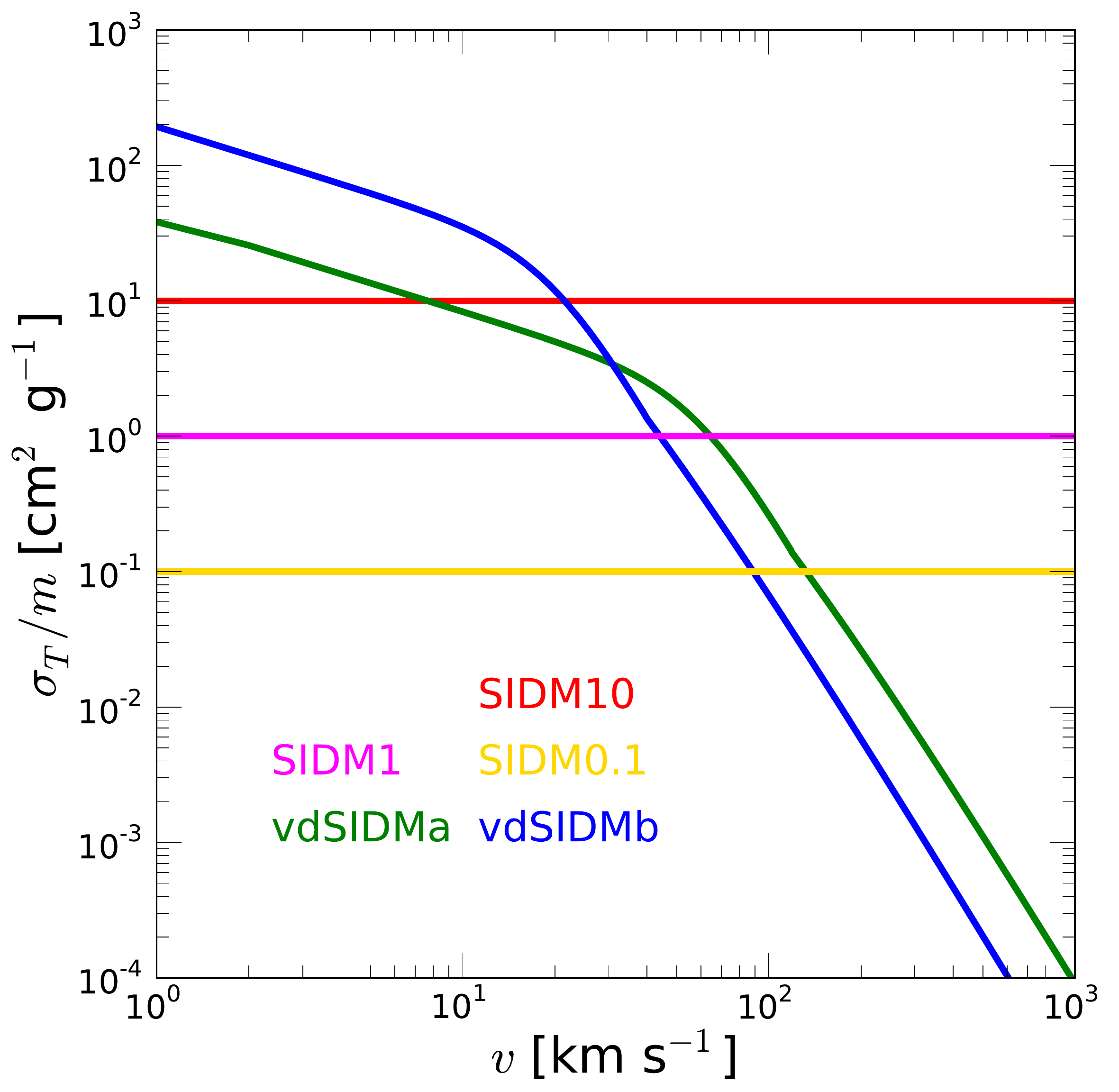}
\caption{Dependence of the momentum-transfer weighted cross-section per unit mass on the relative velocity for the different SIDM models 
considered here. The constant cross section cases with ${\sigma_T/m\!\geq\!1\,{\rm cm}^2\,{\rm g}^{-1}}$ are likely ruled out 
by halo shapes based on X-ray and lensing observations of clusters \citep{Peter2012}. The models with a velocity-dependent 
cross-section are tuned to satisfy all current astrophysical constraints and have been shown to be consistent
with the kinematics of MW dSphs (VZL).}
\label{fig:cross_section}
\end{figure}

Our analysis is based on re-simulations of the Aq-A halo (level 3 resolution) of the Aquarius project \citep{Springel2008}, which
is a set of representative MW-like haloes within the CDM WMAP-1yr cosmology. This halo has a virial mass of
${M_{200}\!\sim\!1.8\times10^{12}\,{\rm M}_{\odot}}$ within a radius of ${r_{\rm 200}\!\sim\!246\,{\rm kpc}}$ (enclosing an average density of 
200 times the critical density). The particle mass in the simulations is ${m_{\rm p}\!\sim\!4.9\times10^4\,{\rm M}_\odot}$ and 
the Plummer equivalent gravitational softening length is ${\epsilon\!\sim\!120\,{\rm pc}}$. We use an algorithm that adds 
dark matter self-scattering to the $N$-body code {\small GADGET-3} for gravitational interactions \citep[last described in][]{Springel2005}.
The algorithm uses a N-body/Monte Carlo approach to represent the microphysical scattering process in the macroscopic context of the
simulation. The connection between this type of approach and the Boltzmann equation is nicely described in Appendix A of \citet{Rocha2012}.
The details of this algorithm can be found in VZL, as well as simple controlled tests that show the agreement between the outcome of the code
and analytical expectations. All dark matter models were simulated starting with the same initial 
conditions and their present-day self-bound subhalo population was identified using the SUBFIND algorithm \citep{Springel2001}.

In addition to CDM, we consider five SIDM cases (recently presented in \citealt{VZ2012} to analyse the impact of self-scattering in
direct detection experiments): three with a constant 
cross section and two with a velocity-dependent one given by a Yukawa-like interaction 
\citep[e.g.][]{Loeb2011}. The transfer cross section scaling with the relative velocity 
can be seen in Fig.~\ref{fig:cross_section}. We note that the formula for ${\sigma_T/m}$ for the velocity-dependent cases
is only valid in the classical regime, once quantum effects are important, the finite interaction length of the Yukawa potential 
cuts off the zero-velocity divergence of the cross section \citep[see e.g.][]{Feng2010}. For our purposes, the quantity of relevance
is ${(\sigma_T/m)~v}$ which goes to zero at zero velocity.
It is clear that for the vdSIDM models, ${\sigma_T/m\!\gg\!0.1\,{\rm cm}^2\,{\rm g}^{-1}}$ at the
characteristic velocities in MW dSphs (the observed velocity dispersion of stars along the 
line of sight is ${\sim\!10\,{\rm km} \,{\rm s}^{-1}}$, e.g. \citealt{Walker2009}). This fact alone already casts a doubt 
on the possibility of SIDM0.1 (${\sigma_T/m\!=\!0.1\,{\rm cm}^2\,{\rm g}^{-1}}$) producing similar results as the vdSIDM cases 
that were shown to be consistent with the kinematics of the MW dSphs in VZL. We note that there is a change in nomenclature
relative to VZL: RefP0$\equiv$CDM, RefP1$\equiv$SIDM10, RefP2-3$\equiv$vdSIDMa-b.

\begin{figure}
\centering
\includegraphics[width=0.475\textwidth]{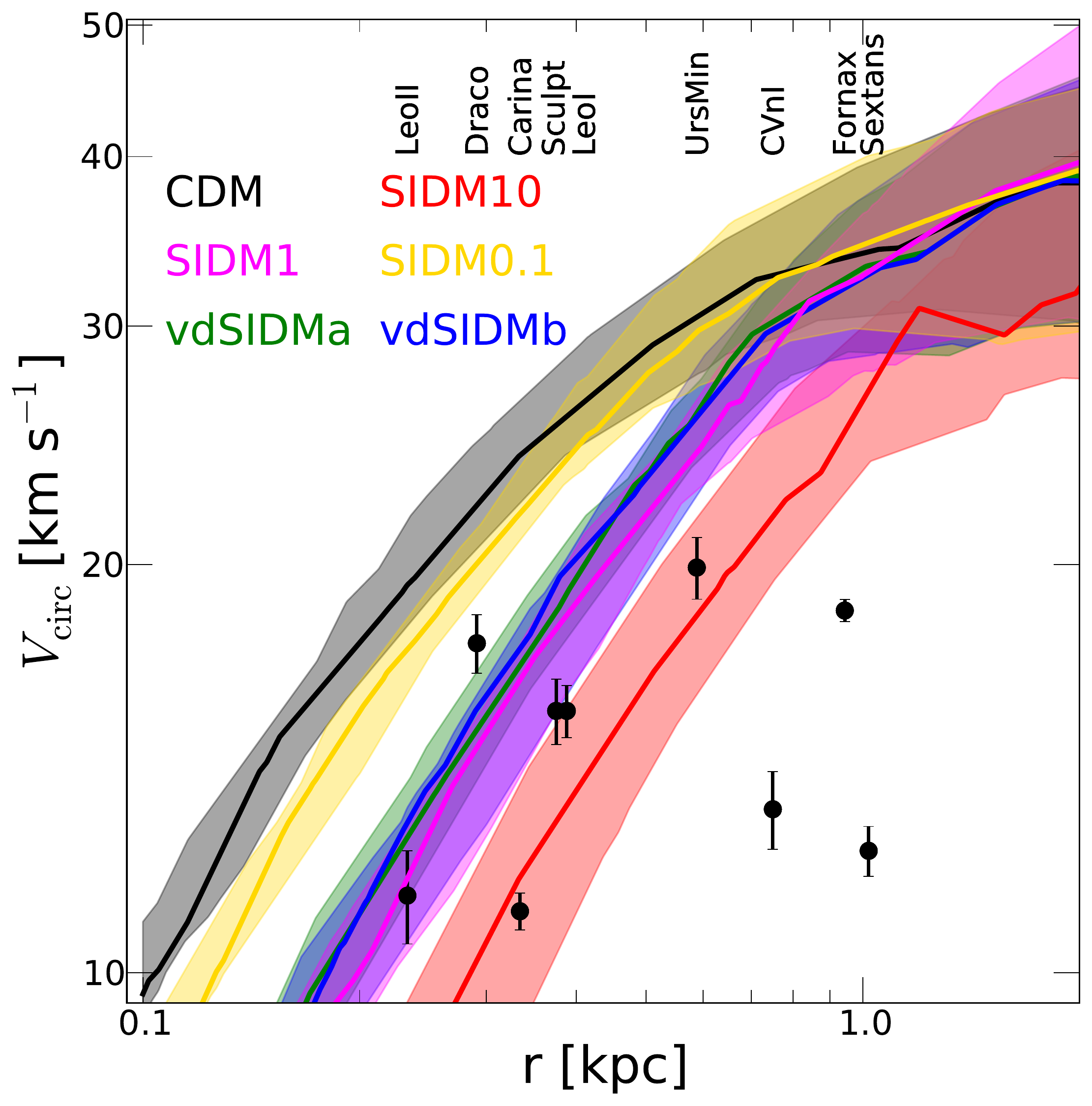}
\caption{The circular velocity profiles at $z = 0$ encompassing the $1^{\rm st}$ and $3^{\rm rd}$ quartiles of the 
distribution of the 15 subhaloes with the largest values of $V_{\rm max}(z=0)$. The symbols with error bars are 
estimates of the circular velocity within the half-light radii for 9 MW dSphs 
\citep{Walker2009,Wolf2010}. Clearly, the most massive CDM subhaloes are inconsistent with the 
kinematics of the MW dSphs. SIDM can alleviate this problem only for a constant scattering cross-section 
${\sigma_T/m\!\gtrsim\! 1\,{\rm cm}^2\,{\rm g}^{-1}}$ (SIDM10 and SIDM1) or if it has a velocity dependence (vdSIDMa and 
vdSIDMb). Current constraints from clusters put an upper limit to the constant cross section case close to 
${\sigma_T/m\!\sim\!0.1\,{\rm cm}^2\,{\rm g}^{-1}}$ (SIDM0.1). This value is too low to solve the too big to fail problem. The 
observational data in the bottom right can be fitted by lower mass subhaloes, not shown here since they are affected 
by the limited resolution of our simulations.}
\label{fig:TBTF}
\end{figure}

Fig.\ref{fig:TBTF} shows the inter-quartile range (i.e., 25-75\%) of the distribution of the 
present-day circular velocity profiles of the 15 subhaloes with the largest values of $V_{\rm max}(z=0)$ 
(the maximum of the circular velocity) within $300\,{\rm kpc}$ halocentric distance. The symbols with error bars
correspond to estimates of the  
circular velocity within the half-light radii of 
the sample of 9 MW dSphs used by \citet{Boylan2011,Boylan2012}. Since current data
for the stars in the dSphs provide an incomplete description of the 6-dimensional phase-space distribution, the 
derived mass profiles are typically degenerate with the velocity anisotropy profile. 
However, the uncertainty in mass that is due to this degeneracy is minimised near the half-light radius, where Jeans models
tend to give the same value of enclosed mass regardless of anisotropy \citep[e.g.][]{Strigari2007,Walker2009,Wolf2010}. 
Observations can then be used to constrain the maximum dark matter density within this radius. CDM clearly predicts a population 
of massive subhaloes that is inconsistent with all the 9 dSphs, whereas for SIDM this problem disappears as long as 
${\sigma_T/m\!\gtrsim\! 1\,{\rm cm}^2\,{\rm g}^{-1}}$ on dSph scales. The currently allowed case with
${\sigma_T/m\!=\!0.1\,{\rm cm}^2\,{\rm g}^{-1}}$ is very close to CDM, only reducing slightly the inner part of 
the subhalo velocity profiles. On the contrary, the vdSIDM models clearly solve the too big to
fail problem. We note that the extent of the too big to fail problem in CDM depends on the mass of the
MW halo, if it is in the low end of current estimates, ${\lesssim\!10^{12}\,{\rm M}_\odot}$, the problem may be resolved 
\citep[e.g.][]{Wang2012}, although a low halo mass may generate other difficulties such as explaining the
presence of the Magellanic Clouds. In the context of SIDM, the lower the mass of the MW halo, the weaker the 
argument against ${\sigma_T/m\!=\!0.1\,{\rm cm}^2\,{\rm g}^{-1}}$.

A simple statistical test of the agreement between the subhalo distributions of two models and the 9 dSphs 
is to compute the chi-square difference associated to the likelihood of having $n^+(n^-)$ data points above (below) the median of the 
distribution of each model. Assuming that the probability distribution of finding $n^\pm$ data points is Poissonian:
\begin{equation}
\Delta\chi^2=2\, ( {\rm ln}(n_1^+!\,\,n_1^-!)-{\rm ln}(n_2^+!\,\,n_2^-!)).  
\end{equation}
Comparing SIDM1 and the vdSIDM models with SIDM0.1, the difference
is driven solely by Draco with the former preferred over the latter with ${\Delta\chi^2\!\sim\!4.4}$ ($2.1\sigma$). Using an interpolation
of our three constant cross section cases, we estimate that ${\sigma_T/m\!\sim\!0.6\,{\rm cm}^2\,{\rm g}^{-1}}$ is the minimum value for which 
$\Delta\chi^2=0$ relative to SIDM1.

\begin{figure}
\centering
\includegraphics[width=0.475\textwidth]{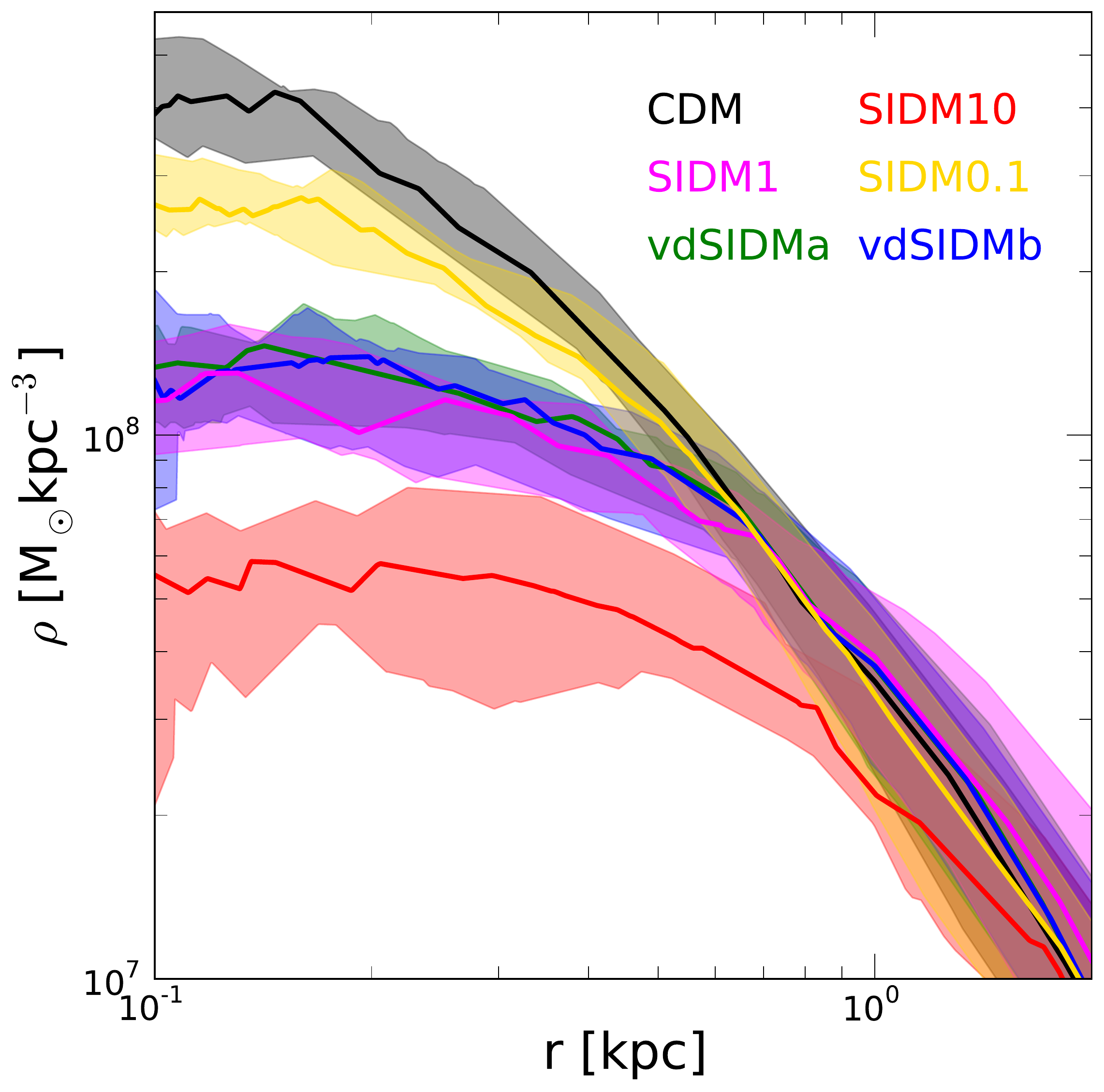}
\caption{Density profile of the 15 subhaloes with the largest $V_{\rm max}(z=0)$ values within CDM and different SIDM
models (see Fig.~\ref{fig:cross_section}). We show the median and $1^{\rm st}$ and $3^{\rm rd}$ quartiles of the subhalo distribution 
for each case. 
The velocity-dependent SIDM cases produce cores of approximately $600\,{\rm pc}$. Of the constant cross section 
SIDM models we explored, the one that is currently allowed by cluster constraints, SIDM0.1 (${\sigma_T/m\!=\!0.1\,{\rm cm}^2\,{\rm g}^{-1}}$), 
only deviates slightly from CDM; the associated core sizes are less than $300\,{\rm pc}$.}
\label{fig:density_profile}
\end{figure}

To show the typical core size and central densities that are predicted by the different SIDM models, we plot
in Fig.~\ref{fig:density_profile} the density profile of the 15 subhaloes with the largest $V_{\rm max}(z=0)$ values. 
A value of ${\sigma_T/m\!\sim\!1\,{\rm cm}^2\,{\rm g}^{-1}}$ is needed for a constant cross section SIDM model to mimic 
the effect of the vdSIDM models and produce ${\sim\!1\,{\rm kpc}}$ cores with central densities of 
${\mathcal{O}(0.1\,{\rm M}_\odot\,{\rm pc}^{-3})}$. If the transfer cross section 
is reduced to ${0.1\,{\rm cm}^2\,{\rm g}^{-1}}$, then the subhaloes are only slightly less dense than in CDM, having
cores (central densities) that are at least twice smaller (higher) than those in the other SIDM cases. 

VZL showed that the SIDM10 and vdSIDM models have convergent density and circular velocity profiles within 
the central density core; we have found the same for SIDM1 and to lesser extent for SIDM0.1. 
Convergence is harder to achieve for CDM since, at a fixed radius, the two-body relaxation 
time is shorter than for SIDM (due to the reduced densities in the latter case). \citet{Power2003} 
showed that the density profile converges at a given radius when the two-body relaxation time is larger than the Hubble 
time at this radius. At the resolution level of our simulations, the convergence radius for CDM is 
$\sim\!600\,{\rm pc}$, which implies that the CDM circular velocity and density profiles shown in Figs.~\ref{fig:TBTF} and 
\ref{fig:density_profile} underestimate the true dark matter content within $\sim\!600\,{\rm pc}$ 
\citep{Springel2008}, whereas for SIDM is at least half of this value. In any case, the expectation is 
that if the density profile of SIDM0.1 has not converged yet, higher resolution would drive it towards higher densities, 
not lower, bringing it even closer to CDM (this is a trend confirmed for the cases analysed in VZL, see their Fig. 9).

By using the fact that some MW dSphs have chemo-dynamically distinct stellar subcomponents that independently trace the same 
gravitational potential, \citet{Walker2011} showed that it is possible to constrain the slopes of their inner mass profiles.
They found that Fornax and Sculptor are
consistent with cored density profiles while cuspy profiles with $\rho\propto r^{-1}$ are ruled out with a significance 
$\gtrsim 96\%$ and $\gtrsim 99\%$, respectively. We use this method to test the consistency
of the SIDM models explored here. 
We found that all SIDM models, except for SIDM0.1, are well fitted by the following three-parameter formula:
\begin{equation}\label{burkert}
\rho(r)=\frac{\rho_0\,r_s^3}{(r+r_c)\,(r^2+r_s^2)},
\end{equation}
which is similar to the Burkert profile \citep{Burkert1995} but with two scale radii $r_s$ and $r_c$. The remaining case, SIDM0.1,
is better fitted by:
\begin{equation}\label{jorge}
\rho(r)=\frac{\rho_0\,r_s^3}{(r+r_c)\,(r+r_s)^2}.
\end{equation}
Using these formulae, we found the best fit parameters for the massive subhaloes in each of the SIDM models. Such fits
are restricted to a radial range between the softening length of our simulations $\!\sim\!120\,{\rm pc}$ and the radius where tidal 
stripping has made the outer logarithmic slope of the density profile steeper than $-3$. The latter restriction is of relevance
only for four subhaloes that are affected significantly by tidal stripping within $\sim\!5\,{\rm kpc}$. 
Two of these are clearly affected within $\sim\!1\,{\rm kpc}$ and should likely be removed in 
a more detailed analysis; they are the least consistent with the data. 
To find the best fit parameters we minimise: 
\begin{equation}\label{goodness}
Q^2=\frac{1}{N_{\rm bins}}\sum_i\left({\rm ln} \rho_i(r_i) - {\rm ln} \rho_{\rm fit}(r_i)\right)^2,
\end{equation}
where the sum goes over all radial bins.Thus defined, $Q$ gives an estimate of the goodness of the fit.
In Table~\ref{table:ref_points}, we give the best fit parameters for the median of the subhalo population 
for each SIDM model (except for SIDM10 which has been ruled out). We note that \citet{Penarrubia2012}
already used Eq.~(\ref{jorge}) to estimate $r_c\gtrsim1\,{\rm kpc}$ for Fornax and Sculptor. Cores of this size are too large
to be consistent with most of the subhaloes in SIDM0.1. 

To test the consistency of the different SIDM models, 
we use the parameters of the fits to compute the slope of the inner mass profile between the pair of 
half light radii (the median likelihood values) of the two distinct stellar subcomponents in Fornax and Sculptor. 
We then test whether this slope is as steep as the lower limit set by the data. The confidence level at which a given
slope is said to be excluded is determined by the fraction of the posterior distribution $f_p$ of allowed slopes that are larger. For 
${\sigma_T/m\!=\!0.1\,{\rm cm}^2\,{\rm g}^{-1}}$, all but 2 subhaloes are excluded at $>95(90)\%$ confidence for Fornax (Sculptor); 
the remaining two subhaloes have values of $f_p\gtrsim0.86(0.81)$ for Fornax (Sculptor). On the contrary, the other SIDM models
(except for SIDM10 that was not analysed) are clearly more consistent with the data with only four subhaloes excluded
at $90\%$ confidence for Fornax (five of the subhaloes actually have $f_p<0.8$), while only three subhaloes are excluded at $>80\%$ confidence
for Sculptor. We found no clear preference between the vdSIDM models and the case with constant
${\sigma_T/m\!=\!1\,{\rm cm}^2\,{\rm g}^{-1}}$. To consider the impact of the non-spherical morphologies of Fornax and Sculptor, we 
repeated the analysis for elliptical rather than circular radii for the stars used to estimate the slope of the mass profiles (see 
sect. 6.1 of \citealt{Walker2011}). We find that for all models Fornax becomes slightly more exclusive while Sculptor is considerably
less exclusive.
 
\begin{table}
\begin{center}
\begin{tabular}{cccc}
\hline
Name  & $\rho(r=200~{\rm pc}) [{\rm M}_\odot\,{\rm kpc}^{-3}]$ & $r_s [{\rm kpc}]$ & $r_c [{\rm kpc}]$ \\
\hline
\hline
vdSIDMa & $1.37\times10^8$    & $0.94$     &   $0.75$                              \\  \hline
vdSIDMb & $1.37\times10^8$    & $0.94$     &   $0.73$                              \\  \hline
SIDM1   & $1.16\times10^8$    & $0.96$     &   $1.33$                              \\  \hline
SIDM0.1 & $2.31\times10^8$    & $0.97$     &   $0.41$                              \\  \hline
\hline
\end{tabular}
\end{center}
\caption{Best fit parameters for the median of the SIDM density profiles of the 15 subhaloes with the largest $V_{\rm max}(z=0)$ values. 
The last two have a constant cross section while the others have a velocity-dependent cross section  
(see Fig.~\ref{fig:cross_section}). 
SIDM1.0 is likely ruled out by cluster observations (see \citealt{Rocha2012}). 
The density profile used for the fits is given by Eq.~(\ref{burkert}) for SIDM1 and the vdSIDM models, and by
Eq.~(\ref{jorge}) for SIDM0.1.}
\label{table:ref_points} 
\end{table}

\section{Discussion and Conclusions}

Self-Interacting Dark Matter (SIDM) offers a promising solution to the dwarf-scale challenges faced by the otherwise-remarkably 
successful Cold Dark Matter (CDM) model. The original idea of a velocity-independent, elastically scattering cross section 
died off quickly, mostly due to the apparently stringent constraint found by \citet{Miralda2002} requiring that the 
cross-section per unit mass was ${\sigma_T/m\!\leq\!0.02\,{\rm cm}^2\,{\rm g}^{-1}}$.
This value is uninteresting, with earlier estimates requiring 
$\sigma_T/m$ to be at least of ${\mathcal{O}(1\,{\rm cm}^2\,{\rm g}^{-1})}$
to create ${\!\sim\!1\,{\rm kpc}}$ cores 
in dwarf-size haloes \citep{Yoshida2000,Dave2001}. \citet{Peter2012} have recently revised earlier constraints on collisional dark matter 
and found them to be overestimated by over an order of magnitude; the current constraint is 
${\sigma_T/m\!\lesssim\!0.1\,{\rm cm}^2\,{\rm g}^{-1}}$. Moreover, these
authors have revived, in a companion paper \citep{Rocha2012}, the velocity-independent SIDM model by suggesting that a value of 
${\sigma_T/m\!=\!0.1\,{\rm cm}^2\,{\rm g}^{-1}}$ is seemingly consistent with the inner structure of the MW dSphs.

Motivated by the prospect of a viable constant cross section SIDM model, 
we investigate the claims from \citet{Rocha2012} using high resolution cosmological SIDM simulations of a MW-size halo. 
Contrary to \citet{Rocha2012}, we are able 
to resolve the sub-${\rm kpc}$ structure of the massive subhalo population to sufficiently small radii for comparison with the
MW dSphs. We find that a velocity-independent SIDM model is consistent with the kinematics of dSphs only if
${\sigma_T/m\!\approx\!1\,{\rm cm}^2\,{\rm g}^{-1}}$ (see Fig.~\ref{fig:TBTF}), i.e., a value of this order is required to solve the {\it too
big to fail problem} \citep{Boylan2011,Boylan2012}. If the cross section is lower by an order of magnitude, the subhalo population is still too
dense to be consistent with the MW dSphs. On the other hand, as shown already in VZL, velocity-dependent SIDM models with
a Yukawa-like interaction (as proposed in \citealt{Loeb2011}, see Fig.~\ref{fig:cross_section}) successfully solve the too big to fail problem.

\begin{figure}
\centering
\includegraphics[width=0.475\textwidth]{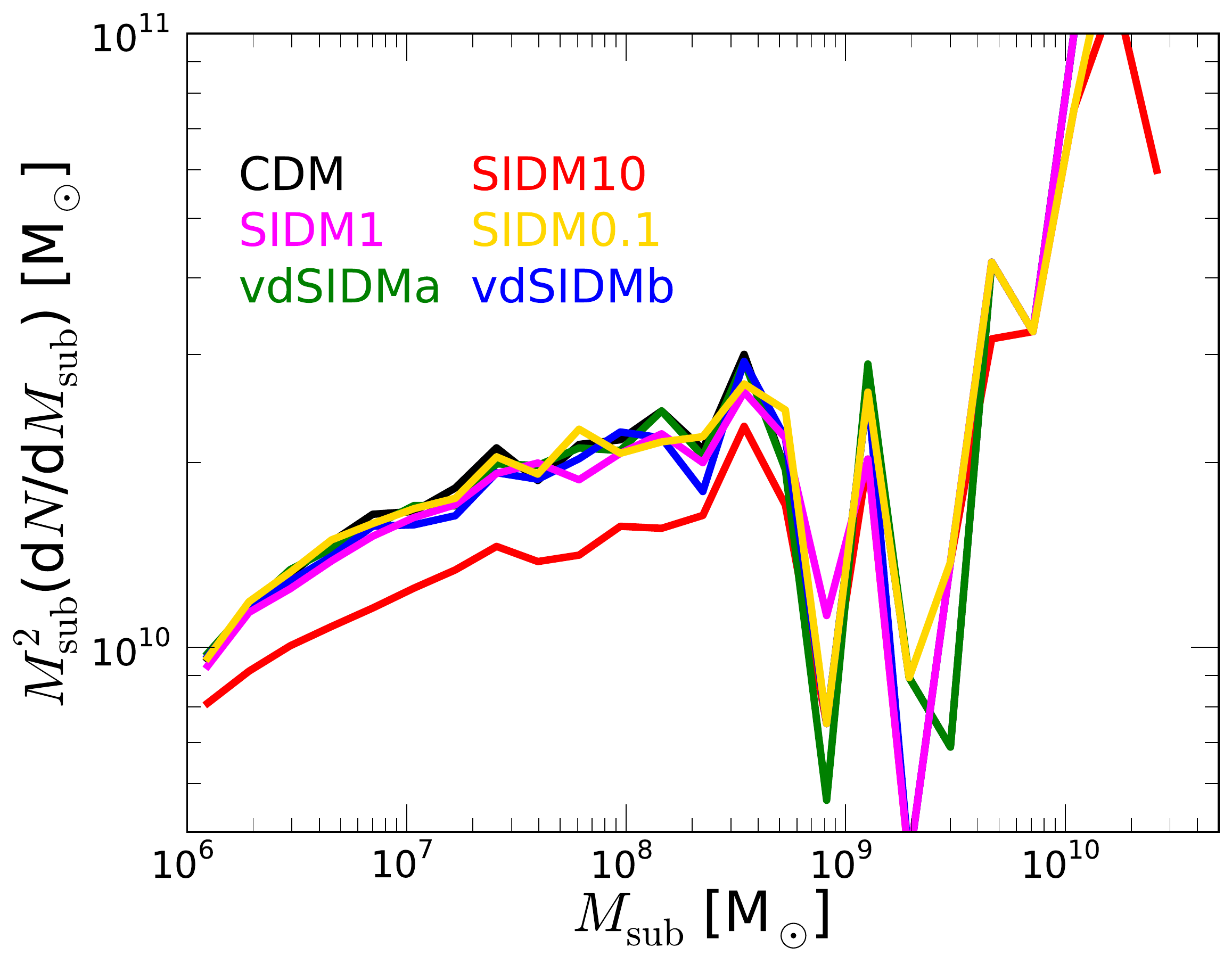}
\caption{Subhalo mass function for a MW-size halo within CDM and different elastic SIDM models. The only model
that leads to a difference relative to CDM has a constant cross section of ${\sigma_T/m\!=\!10\,{\rm cm}^2\,{\rm g}^{-1}}$, 
which is clearly ruled out by cluster observations.}
\label{fig:mass_function}
\end{figure}

We also use the inner slopes of the mass profiles of Fornax and Sculptor, from the analysis of \citet{Walker2011},
as examples to test the consistency of the different models we simulate here. For a velocity-independent SIDM model with 
${\sigma_T/m\!\sim\!0.1\,{\rm cm}^2\,{\rm g}^{-1}}$, we find that 13 of of the 15 subhaloes with the largest $V_{\rm max}(z=0)$ values 
are inconsistent with the data from
Fornax (Sculptor) at $>95(90)\%$ confidence (the other two are inconsistent at $\gtrsim81\%$ confidence). A constant cross section 
ten times larger is as consistent as the velocity-dependent SIDM models explored here with only four (three) of the top 15 subhaloes 
excluded at $>90(80)\%$ confidence in the case of Fornax (Sculptor);  
for all these cases, there are several subhaloes that are unambiguously consistent with the data. 

According to the analysis of \citet{Peter2012}, a constant cross section of ${\sigma_T/m\!=\!1\,{\rm cm}^2\,{\rm g}^{-1}}$
is likely inconsistent with the observed halo shapes of several clusters. We have now shown that ${\sigma_T/m\!=\!0.1\,{\rm cm}^2\,{\rm g}^{-1}}$ 
is too close to CDM to represent a distinct alternative. An interpolation of our simulations suggests that the central densities 
of the massive subhaloes would be consistent with the MW dSphs if ${\sigma_T/m\!\sim\!0.6\,{\rm cm}^2\,{\rm g}^{-1}}$. We conclude 
that the hypothesis of a
constant scattering cross section as solution to the core-cusp problem remains viable but within a very narrow range of $\sigma_T/m$ values. 
The challenges to make a definitive test of this hypothesis are twofold: the cluster-constraints need be refined, and 
the impact of conservative baryonic processes needs to be estimated. Although adding gas physics is the next step of SIDM simulations,
a challenge to make SIDM an even more attractive alternative to CDM is the prospect of explaining
the observed scarcity of MW satellites and field dwarfs \citep[e.g.][]{Klypin1999,Zavala2009} without invoking extreme
baryonic processes. As we show in Fig.~\ref{fig:mass_function}, all allowed {\it elastic} SIDM
models essentially produce the same abundance of dwarf-size haloes as in CDM. A promising possibility is that of exothermic interactions 
between excited and non-excited states of dark matter \citep[e.g.][]{Loeb2011}. The velocity kick imparted during the collision
might be large enough to cause the evaporation of low-mass haloes. 

\section*{Acknowledgements}

We thank Volker Springel for giving us access to {\sm GADGET-3},  Niayesh Afshordi, 
Manoj Kaplinghat and Paul Steinhardt for fruitful discussions, Mike Boylan-Kolchin, 
Fabio Governato, Lars Hernquist, Avi Loeb, Jorge Pe\~narrubia and Simon White for useful suggestions. 
JZ is supported by the University of Waterloo and the Perimeter Institute for
Theoretical Physics. Research at Perimeter Institute is supported by the
Government of Canada through Industry Canada and by the Province of Ontario
through the Ministry of Research \& Innovation. JZ acknowledges financial
support by a CITA National Fellowship. M.G.W. is currently supported by NASA
through Hubble Fellowship grant HST-HF-51283.01-A, awarded by the Space
Telescope Science Institute, which is operated by the Association of
Universities for Research in Astronomy, Inc., for NASA, under contract
NAS5-26555. MV acknowledges support from NASA through Hubble Fellowship grant
HST-HF-51317.01. 

\bibliography{paper}

\label{lastpage}

\end{document}